\documentclass[twocolumn,preprintnumbers,prl,floatfix,superscriptaddress,amsmath,amssymb]{revtex4}
\usepackage{epsfig}
\usepackage{subfigure}
\usepackage{amsmath}
\usepackage{amssymb}
\usepackage[usenames]{color}
\usepackage{setspace}
\usepackage{times}
\usepackage{graphicx}
\usepackage{dcolumn}
\usepackage{bm}
\input epsf

\begin{document}

\author{Juan G. Restrepo}
\affiliation{Department of Physics and Center for Interdisciplinary Research on Complex Systems, Northeastern University, Boston, MA, 02115 USA}

\author{Alain Karma}
\affiliation{Department of Physics and Center for Interdisciplinary Research on Complex Systems, Northeastern University, Boston, MA, 02115 USA}

\date{\today}

\begin{abstract}
Spiral wave propagation in period-2 excitable media is often accompanied by
line-defects, the locus of points with period-1 oscillations.
Here we investigate spiral
line-defects in cardiac tissue where period-2 behavior has a
known arrhythmogenic role. We find that the number of line defects, which
is constrained to be an odd integer, is three for a freely rotating spiral, with and without meander,
but one for a spiral anchored around a fixed heterogeneity.
We interpret analytically this finding using a simple theory where spiral wave
unstable modes with different numbers of
line-defects correspond to quantized solutions of a Helmholtz equation. Furthermore,
the slow inward rotation of spiral line-defects is described in different regimes.


\end{abstract}

\title{ Line-Defect Patterns of Unstable Spiral Waves in Cardiac Tissue}


\pacs{}

\maketitle

Spiral waves are observed in extremely diverse physical and biological
excitable media and are known to play a key role in the genesis of
abnormally rapid life-threatening heart rhythm disorders \cite{weiss}.
Despite considerable progress to date,
complex spatiotemporal behaviors resulting from
unstable spiral wave propagation remain poorly understood theoretically with the
exception of meander \cite{barkley,hakim}, a classic spiral
core instability
with flower-like tip trajectories.
A particularly rich dynamics results from instabilities in
period-2 media where the local dynamics of the medium,
i.e. the dynamics of uncoupled excitable elements,
exhibits a period-doubling bifurcation as a function of parameters
of the medium or the external stimulation frequency.
Although period-2
behavior has been seen in different excitable and oscillatory media, it has
received particular attention in a cardiac context.
The hallmark of period-2 behavior in this context is alternans, a beat-to-beat
alternation in the duration of cardiac excitation, which has been linked
to the onset of lethal heart rhythm disorders \cite{karmapt}.

Unstable spiral wave propagation in period-2 media is
invariably accompanied by  ``line-defects'',
which are the locus of points where the
dynamics is locally period-1. Line-defects are generally present in these media when
plane waves radiating out of the core region are unstable at the spiral
rotation period, independently
of whether meander is present or not.
Studies in {\it in vitro} cardiac cell tissue cultures \cite{leepnas,kimpnas},
chemical reactions \cite{parklee,parklee2,marts}, and coupled oscillators \cite{goryachev,wu} have
revealed the existence of a rich variety of patterns ranging from one and three line-defect structures \cite{parklee2},
to phase bubbles \cite{parkleeprl2008}, to line-defect turbulence \cite{parkleeprl1999}.
Spiral wave breakup
in models of cardiac excitation has also been found in parameter
regimes of local period-2 dynamics, and hypothesized
in this context as a potential mechanism for heart fibrillation \cite{weiss,karmaprl,karmachaos,fenton}.
Spiral line-defect patterns, however, have not been systematically
investigated in cardiac tissue.


In this Letter, we investigate the selection and
dynamics of line-defect patterns resulting from unstable spiral
wave propagation in cardiac tissue. Moreover, we interpret our findings using an amplitude equation
framework recently used to study the evolution of line-defects during periodic
stimulation from a single site \cite{blas}. In this framework, the spatiotemporal modulation of the phase and amplitude
of period-2 oscillations is described by a simple partial differential equation that can be
readily analyzed.
Our study is based on the
standard wave equation for cardiac tissue
\begin{equation}
\partial_t V = \gamma \nabla^2 V - I_m(V,\vec y)/C_m, \label{rd1}
\end{equation}
where $V$ is the transmembrane voltage, $\gamma$ is the voltage diffusion coefficient, $C_m$ is the membrane capacitance,
and $\vec y$ is a vector of gate variables that controls the flow of ions through the membrane, and hence
the total membrane ionic current $I_m$. We studied different models of
$I_m(V,\vec y)$ and gating kinetics
to explore universal features of line-defect patterns that depend on
qualitative properties of core and plane wave instabilities. The latter
are manifested either as
{\it stationary} \cite{blas,watanabe} or {\it traveling} \cite{blas}
spatial modulations of period-2 oscillation amplitude with
an intrinsic spatial scale determined by parameters of the excitable medium \cite{blas}.
These spatial modulations have nodes with period-1 dynamics
in one dimension, or nodal lines in two, which correspond here to
line-defects in the spiral far-field. We therefore chose models
to explore line-defect patterns for stationary and
traveling nodes with and without
meander. The model of Ref.~\cite{karmachaos} has pinwheel spirals (no meander)
and stationary nodes under periodic pacing. The other
two models of Ref. \cite{blas3v}
and Ref. \cite{blas} both exhibit meander and have fixed
and traveling nodes, respectively.

\begin{figure}
\centering \epsfig{file = 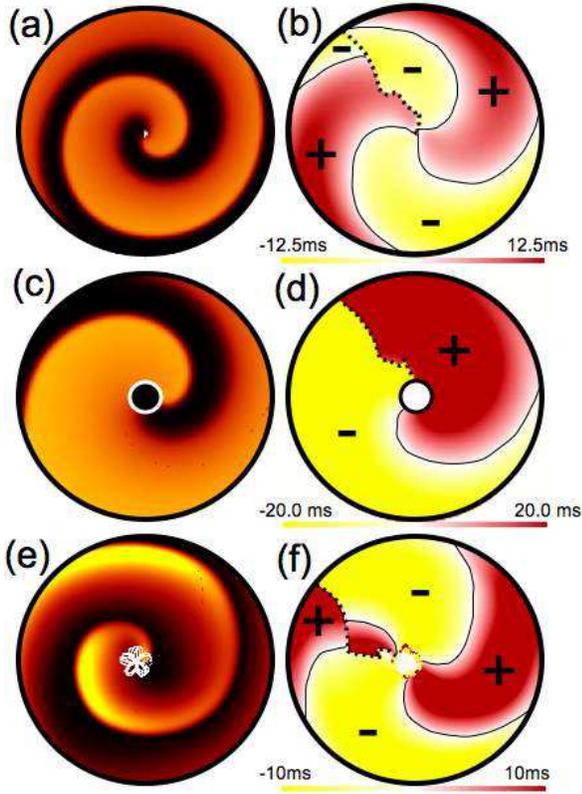, clip =
,width=0.90\linewidth }
\addtolength{\belowcaptionskip}{-0.7cm}\addtolength{\abovecaptionskip}{-0.3cm}
\caption{Membrane voltage (left) and corresponding alternans
amplitude $a$ (right) for different ionic models and geometries:
model of Ref. \cite{karmachaos} without (a and b; $Re=1.15$) and
with a circular inexcitable disk of radius $r_i=0.5$~cm (c and d;
$Re=1.25$), and model of Ref. \cite{blas3v}  without disk (e and f;
$t_{f2} = 30$ ms, $t_{si} = 10$ ms, $t_{h2}=2.0$ ms). In panels (a),
(c), and (e), grey (orange online) indicates depolarization, and
white indicates the trajectory of the spiral tip.  In panels (b),
(d), and (f), dark grey (red online) and light grey (yellow online)
correspond to $a> 0$ and $a< 0$ regions, separated by line-defects
(solid lines); the grey (color online) scale does not vary outside
the indicated range of $a$. Line-defects in (b), (d), and (f),
rotate counterclockwise with a period that is approximately $17$,
$8$, and $9$ times the clockwise spiral rotation period in (a), (c),
and (e), respectively.
The discontinuity of $a$ across the dashed line is a consequence
of the definition of the common beat number (see text).} \label{spiral}
\end{figure}

Freely propagating spiral waves
in all three models were studied by numerically solving
Eq.~(\ref{rd1}) in a circular domain of  radius $r_e = 3$ cm with
no-flux boundary condition, $\partial_r V|_{r=r_e}=0$.
Anchored spirals were studied by introducing an inexcitable
disk of radius $r_i$ and imposing no-flux
conditions on both the inner and outer radii, $\partial_r V|_{r=r_i}=\partial_r V|_{r=r_e}=0$.
We implemented the phase-field method of Ref.
\cite{fentonetalchaos}
that automatically handles no-flux boundary conditions
in an arbitrary geometry using a finite-difference representation
of the Laplacian on a square grid,
and iterated Eq.~(\ref{rd1}) using a simple
explicit Euler scheme.
Model parameters are identical
to the published ones except those listed in Fig. 1. The latter
were chosen for intermediate action potential
duration restitution slopes, which suffice to
produce unstable spiral waves with line-defects
in each geometry, but are not steep enough to cause wave breakup
in this domain size.

We used a half plane wave as initial condition to initiate
a spiral wave (obtained
by first triggering a full plane wave and
resetting part of the circular domain to the resting state).
To track line defects, we define
at each point ${\bf x}$ and time $t$ a local beat number $n({\bf x},t)$,
set everywhere initially to zero after the half plane wave is created,
and increased by one at the end of each action potential, i.e. every time that the voltage
$V({\bf x},t)$ crosses a fixed threshold $V_c$ with $dV/dt<0$.  We then define
the period-2 alternans amplitude as
\begin{equation}
a({\bf x} ,t) =(-1)^{n_c(t)} \left[D({\bf x},n_c(t)) - D({\bf x},n_c(t)-1)\right]/2 \label{adef}
\end{equation}
where $D({\bf x},n)=\int_{V({\bf x},t')>V_c,n({\bf x},t') = n-1}dt'$
is the local action potential duration (APD) and  $n_c(t) \equiv
\min_{{\bf x}}n({\bf x},t)$ is the common beat, i.e. the largest
beat number that has been registered at all points at time $t$. The
line-defects are then the locus of points where $a({\bf x} ,t)=0$ at
any instant of time. The use of a common beat number introduces here
a discontinuity in $a$ (indicated by dashed lines in Fig.
\ref{spiral}) since the APD of a given beat might change as the wave
front rotates around the spiral tip. This discontinuity, however,
does not affect the dynamics. Other methods to track line defects
\cite{parklee,zhan} yield similar results except for unessential
imaging differences.


Results of simulations that pertain to the selection of the number
of line-defects are shown in Fig.~\ref{spiral}. The top four panels reveal that the
pinwheel spirals simulated with
the two-variable model of Ref. \cite{karmachaos} exhibit three line-defects
when propagating freely in spatially homogeneous tissue,
but only one line-defect when anchored around
an inexcitable disk of $0.5$ cm radius. Furthermore, the bottom two panels show that for
the more physiologically realistic three-variable model of Ref. \cite{blas3v}, freely
propagating spirals still exhibit three-line defects even though the spiral
tip meanders.

\begin{figure}[t]
\centering \epsfig{file =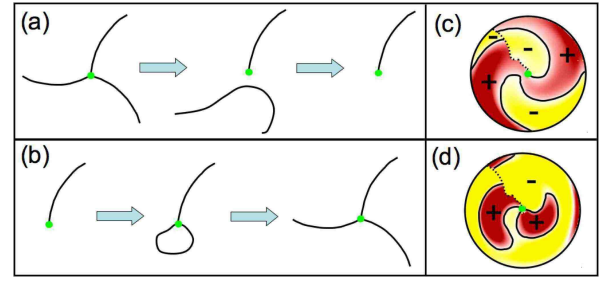, clip = ,width=0.9\linewidth
}
\addtolength{\belowcaptionskip}{-0.7cm}\addtolength{\abovecaptionskip}{-0.2cm}
\caption{Schematic illustration of transitions from (a) three to one
and (b) one to three line-defects and corresponding examples from
simulations of the model of Ref. \cite{karmachaos} with the same
conventions and parameters as in Fig.~\ref{spiral} (b). The small
circle (green online) in (a) and (b) represents the spiral core
region or a small anchoring obstacle.} \label{schematic}
\end{figure}

Since anchored spirals become free in the limit of vanishing obstacle size,
one would expect transitions from one to three (three to one)
line-defects to occur with decreasing (increasing) obstacle size.
Indeed, for the model of Ref.~\cite{karmachaos},  we found three
line defects for obstacles with diameter smaller than $\sim0.1$ cm, including the freely propagating
pinwheel spiral ($r_i=0$) in Fig.~\ref{spiral} (b), and one line defect for diameters larger than $\sim0.3$ cm as
in the example of Fig.~\ref{spiral} (d). For intermediate diameters, we found complex behaviors
marked by transitions from three to one or one to three line-defects. The former
occur when two line defects merge into one line defect that moves away from the core, and the
latter when a phase bubble enclosed by a line-defect loop nucleates in, and expands from, the
core, as illustrated in Fig.~\ref{schematic}. We find the same qualitative behavior in
the model of Ref. \cite{blas3v} except that meander makes the transitions between patterns
with different numbers of line-defects more complex.

Let us now turn to interpret our results
in the amplitude equation framework \cite{blas}.
For simplicity, we restrict our analysis to non-meandering spiral
waves. Furthermore, to keep the analysis tractable, we
first assume that the propagation wave speed is constant and
relax this assumption subsequently when examining the motion
of line defects.
With this assumption, linear perturbations of a steady-state rigidly rotating
spiral wave with period $T$ obey the equation
\begin{equation}
T \partial_t a= \sigma a + \xi^2 \nabla^2a, \label{linear}
\end{equation}
where $a$ is the alternans amplitude
subject to the radial  $\partial_r a|_{r_i}=\partial a|_{r_e}=0$ and
angular $a(\theta+2\pi,t) = -a(\theta,t)$ boundary conditions. The latter constrains the
number of line defects to be an odd integer and results
from the change in beat number across any closed circuit enclosing the spiral tip
for steady-state alternans.
It follows directly from the definition of $a$ [Eq.~(\ref{adef})] and the
requirement that the voltage be continuous everywhere in space.
In addition, $\sigma=\ln f'$, where $f'$, the slope of the action potential duration restitution
curve defined by $D^{n+1} = f(T-D^n)$, controls the onset of alternans and
$\xi\sim (\gamma D)^{1/2}$, where $D$ is the value of
the action potential duration at the period-doubling bifurcation ($\sigma=0$),
measures the scale over which the voltage dynamics is diffusively
coupled on the time scale of one beat.

\begin{figure}
\centering \epsfig{file =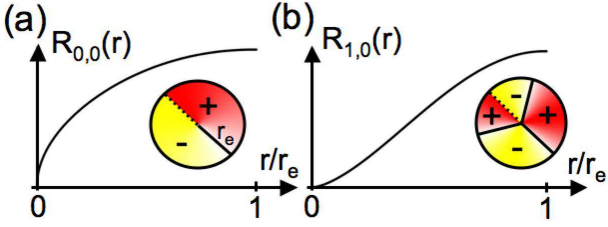, clip = ,width=0.95\linewidth
}
\addtolength{\belowcaptionskip}{-0.7cm}\addtolength{\abovecaptionskip}{-0.5cm}
\caption{Theoretical radial and angular (insets) dependence of the
alternans amplitude for (a) $1$ and (b) $3$ line defect modes.}
\label{tentative}
\end{figure}

This linear stability problem is easily solved by
the substitution $a({\bf r},t) \sim e^{\Omega t} \Psi(r,\theta)$  that transforms
Eq. (\ref{linear}) into a Helmholtz equation for $\Psi(r,\theta)$. The
latter can then be solved by separation
of variables with the substitution $\Psi(r,\theta)\sim R(r)\Theta(\theta)$. The angular part is found to be
$\Theta_n(\theta) =  \sin\left((n+1/2) \theta\right)$, where mode $n$ corresponds to $2n+1$ line-defects.
The radial part obeys a Bessel equation. For $r_i>0$, it has solutions
$R_{n,m}(r) \propto J'_{-n-1/2}(k_{n,m} r_e)J_{n+1/2}(k_{n,m} r) - J'_{n+1/2}(k_{n,m} r_e)J_{-n-1/2}(k_{n,m} r)$
that satisfy the outer radial boundary condition $\partial a|_{r_e}=0$, where $n,m = 0,1,\dots$, and the inner
condition $\partial_r a|_{r_i}=0$ determines $k_{n,m}$, and hence the growth rate
$\Omega_{n,m} T = \sigma - \xi r_e^{-2} k^{2}_{n,m}$.
We find that the smallest $k_{n,m}$ occurs for $n=0$ independently of the ratio $r_e/r_i$. Therefore, the mode corresponding to a single line defect is the most unstable when the spiral is anchored.
This agrees with our numerical observations in Fig.~\ref{spiral} (d).
For freely rotating spirals, $r_i = 0$, $J_{-n-1/2}(r)$ diverges at the origin, so the solutions are
$R_{n,m}(r) \propto J_{n+1/2}(k_{n,m} r)$, where $k_{nm} r_e$ is the $m^{th}$ zero of $J'_n(r)$.
The most unstable modes are $n=0$ and $n=1$ corresponding to $1$ and $3$ line defects, respectively (see Fig.~\ref{tentative}). However, $J_{1/2}(k_{0,0} r)$ has a divergent derivative that is incompatible with the physical
requirement that the voltage, and hence the APD, must vary smoothly on a scale $\xi$. On the other hand, $J_{3/2}(k_{1,0} r)$ smoothly vanishes at the origin.  Therefore,
in this case, the boundary condition at the origin selects a 3-line-defect pattern
as observed in Fig.~\ref{spiral} (b).
Interestingly, a 3-line-defect pattern is also
selected with meander present [Fig.~\ref{spiral}~(f)], thereby suggesting
that the boundary condition on $a$ on the outer scale of the line-defect pattern
is not strongly affected by meander.



\begin{figure}
\centering \epsfig{file =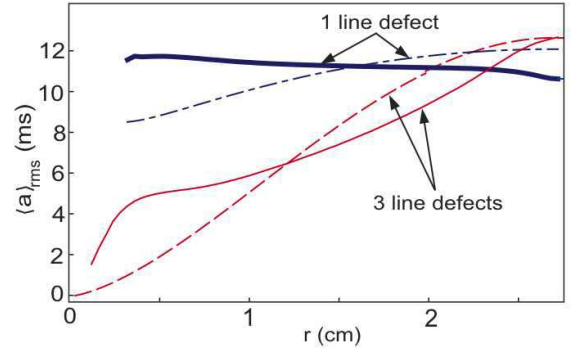, clip = ,width=0.90\linewidth
}
\addtolength{\belowcaptionskip}{-0.80cm}\addtolength{\abovecaptionskip}{-0.4cm}
\caption{$\langle a\rangle_{rms}$ versus radial distance $r$ for the
pinwheel spiral (thin solid line, red online) and anchored spiral
(thick solid line, blue online), and respective theoretical radial
modes $R_{1,0}(r)$ (thin dashed, red online), and $R_{0,0}(r)$
(dotted-dashed line, blue online).} \label{raial}
\end{figure}

The analysis also predicts qualitative features of the radial
distribution of alternans amplitude for three- and one-line-defect
patterns of Figs. \ref{spiral}(b) and \ref{spiral}(d), respectively. Fig.~\ref{raial} compares
the numerical radial distributions
of root-mean-square amplitude $\langle a \rangle_{rms}$ averaged over a full line-defect rotation period for three
line-defects (thin solid line) and one line-defect (thick line), with
the corresponding radial modes from the theoretical analysis (dashed lines)
scaled to have the same radial average as the observed curves. The theory predicts well that the
alternans amplitude is more strongly suppressed near the core for the larger number
of line defects.

So far our analysis has assumed that the wave speed is constant, which predicts that
line-defects extend straight out of the core and are stationary,
as implied by the angular distribution $\sin\left((n+1/2) \theta\right)$
of linearly unstable modes (see Fig.~\ref{tentative}). In contrast, simulations in Fig. \ref{spiral} show that line-defects
have a spiral shape and slowly rotate inward in the opposite direction of
the spiral wavefront. Line-defect motion can generally be induced both by
line-defect curvature and the dependence of the wave speed $c$ on the interval $I$
between two waves, known as the conduction velocity (CV) restitution curve
in the cardiac literature. While a full stability analysis that includes these effects
would be required to treat line-defect motion in general, two important limiting
cases can be readily analyzed.

\begin{figure}
\centering \epsfig{file = 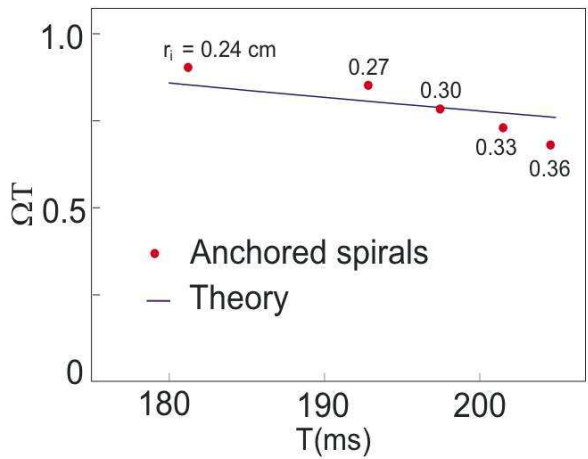, clip =
,width=0.95\linewidth }
\addtolength{\belowcaptionskip}{-0.8cm}\addtolength{\abovecaptionskip}{-0.5cm}
\caption{Comparison of numerical (circles, red online) and
theoretical (solid line) line-defect rotation frequencies for spiral
waves anchored to disks of varying radii for the model of Ref.
\cite{karmachaos}.} \label{frecues}
\end{figure}

The first pertains to anchored spiral waves
for medium parameters where plane waves
paced at the spiral rotation period exhibits stationary line-defects,
as for the model of Ref. \cite{karmachaos}
studied here. In this case, we expect line-defect motion to
be generated predominantly by the spiral wavefront dynamics
around the anchoring obstacle.
Neglecting wavefront curvature effects,  this dynamics should be
approximately described by that of a propagating pulse
in a one-dimensional ring of perimeter  $L=2\pi r_i$ \cite{blas,court}.
To test this hypothesis, we computed the quasiperiodic frequency $\Omega$ of the
local medium dynamics induced by
line-defect rotation for anchored spirals for the model of Ref. \cite{karmachaos}. The frequency
was obtained by fitting the time series $a({\bf r},jT)/a({\bf r},0)$ at a single point ${\bf r}$ to $\eta^j\cos(\Omega T j+\delta)$, with $\eta$, $\Omega$, and $\delta$ the fitting parameters. For the theory,
we used the dispersion relation giving the quasiperiodic frequency
$\Omega$ modulating alternans, $a \propto e^{i\Omega jT}$, in a one dimensional ring derived in Ref.~\cite{blas}
\begin{equation}
e^{i\Omega T}\left( 1-\frac{i}{2\Lambda k} \right) = (1-i w k -\xi^2k^2) f'(I) +\frac{i}{2\Lambda k},
\end{equation}
where $k = \pi/L + \Omega T/L$ is the wavenumber corresponding to a single line defect and $\Lambda = c'(I)/(2c^2)$. The APD- and CV-restitution
curves, $f(I)$ and $c(I)$, were calculated in a one-dimensional cable as in
Ref. \cite{blas}. In addition, the intercellular coupling parameters $\omega$ and $\xi$ were estimated as $\omega \sim 2\gamma/c$ and $\xi \sim (\gamma D)^{1/2}$ \cite{blas}. The comparison in Fig.~\ref{frecues} shows
that the ring-based theory predicts reasonably well the frequency of
line-defect rotation for anchored spiral waves
of different period $T$, which was varied here by increasing the obstacle
radius $r_i$ in the simulations. 

The opposite limit that can also be readily understood is the one
where plane waves paced at the spiral rotation period exhibit line-defects that
move towards the pacing site, which generally occurs
for steeper CV-restitution. In this case, line-defect motion
is expected to be dominated by the far-field spiral dynamics \cite{blas}.
We have checked that, for the two-variable model
of Ref.~\cite{blas}, spiral line-defects indeed rotate inward
with a frequency equal to the product of the velocity of the
planar line-defects and the inverse of their spacing. This property was purposely
checked in a domain much larger than the spiral wavelength
($r_e=18$ cm) and with an obstacle size ($r_i=0.72$ cm)
sufficient to prevent spiral wave breakup inherent in this model. However, we
expect this behavior to be generic for systems with traveling planar line-defects
and to also apply to freely rotating spirals with three line defects for parameters
where breakup does not occur.

In summary, we have surveyed spiral line-defect patterns in
simplified models of cardiac excitation with period-2 dynamics.
Although far from exhaustive, this survey yields the striking
finding that freely propagating
and anchored spiral waves select different
numbers of line-defects. This opens up the possibility to
distinguish free and anchored spiral waves in cardiac tissue by
monitoring the number of line-defects. We have shown that spiral wave
unstable modes with different numbers of
line-defects correspond to topologically
quantized solutions of a Helmholtz equation.
In this framework, the boundary condition on the period-2 oscillation amplitude in the
spiral core, which is fundamentally different for free
and anchored spirals, is responsible for selecting
the number of line defects. Furthermore, we have found that
spiral line-defect inward rotation can be driven
either by the core or far-field wavefront dynamics,
with concomitantly different frequencies.
Our results suggest that the observation of single-line-defect spirals in cardiac tissue culture \cite{leepnas,kimpnas} may be
a consequence of anchoring on small millimeter-size heterogeneities. However, the dynamics
in real tissue is also influenced by the coupling of voltage
and intracellular calcium dynamics \cite{weiss,karmapt},
which has been neglected here. The investigation of the effect of
this coupling on line-defect dynamics and its relationship
to wave breakup is an interesting future project.
Finally, the previous finding of free spirals with one line-defect
\cite{zhan} in period-2 media with qualitatively different excitable
dynamics than cardiac tissue suggests that other pattern selection
mechanisms may be operative in different media. These differences
also remain to be elucidated. 

We thank Blas Echebarria for valuable discussions. This work was supported by NIH Grant  No.~P01 HL078931.

\begin{spacing}{1.0}
\bibliographystyle{plain}

\end{spacing}

\end{document}